\documentclass[%
 reprint,
 superscriptaddress,
 amsmath,amssymb,
aps,
prl
]{revtex4-1}

\usepackage{comment, color}
\usepackage{braket}

\usepackage{graphicx}
\usepackage{dcolumn}
\usepackage{bm}

\begin{document}

\title{Second-order topological non-Hermitian skin effects}

\author{Ryo Okugawa}
 \affiliation{%
 	Graduate School of Information Sciences, Tohoku University, Sendai 980-8579, Japan
 }%
\author{Ryo Takahashi}
 \affiliation{%
	Department of Physics, Tokyo Institute of Technology, 2-12-1 Ookayama, Meguro-ku, Tokyo, 152-8551, Japan
}%

\author{Kazuki Yokomizo}
 \affiliation{%
 	Department of Physics, Tokyo Institute of Technology, 2-12-1 Ookayama, Meguro-ku, Tokyo, 152-8551, Japan
 }%

\date{\today}

\begin{abstract}
Higher-order topology realizes topologically robust corner modes as a manifestation of nontriviality. 
We theoretically propose non-Hermitian skin effects which stem from second-order topology of chiral-symmetric Hermitian systems.
It is found that the skin modes are localized at the corners.
We demonstrate two types of second-order topological skin effects
by two-dimensional intrinsic and extrinsic second-order topology.
The intrinsic second-order topological skin effect is characterized topologically by bulk inversion symmetry as well as chiral symmetry.
Meanwhile, the extrinsic second-order topological skin effect occurs from the topological correspondence between the edges and corners.
We show that the non-Hermitian skin modes emerge by using a relationship between second-order and conventional first-order topology.
\end{abstract}

\maketitle

\textit{Introduction.}
Since the discovery of the quantum spin Hall insulator \cite{Kane05, Kane05L2, Fu06, Fu07},
many topological phases have been studied theoretically and experimentally \cite{Hasan10, Qi11, Ando15, Chiu16}.
The concept of topological phases has been recently generalized to higher-order topology and non-Hermitian topology.
In $d$-dimensional $n$th-order topological phases,
topologically protected states appear at the $(d-n)$-dimensional boundaries such as corners
\cite{Benalcazar17S, Benalcazar17B, Hashimoto17, Song17, Slager15, Langbehn17, Schindler18, Khalaf18X, Khalaf18B, Ahn18, Matsugatani18, 
	Lin18, Fukui18, Geier18, Luka19, Hayashi18, Hayashi19, Okugawa19B, Ahn19, Wang19L, Park19, Lee20, Tanaka20, Takahashi20,
	Liu19L, Edvardsson19, Zhang19L, Luo19, Ezawa19B2}.
Meanwhile, non-Hermiticity enriches topological phenomena which do not emerge in Hermitian systems
\cite{Martinez18, Ghatak19}.
Such examples are exceptional points \cite{Szameit11, Ramezani12, Zhen15, Cerjan16, Xu17, 
	Shen18, Kozii17a, Cerjan18, Yoshida18, Zyuzin18, Carlstrom18, Okugawa19BR, Budich19, Rui19, Zhou19Opt, Yoshida19BR,
	Yang19, Wang19, Moors19, Zyuzin19, Zhang19, Cerjan19, Carlstrom19, Yoshida19B, Kawabata19L, Kimura19, Yokomizo20a}
and skin effects \cite{Gong18, Yao18L1, Yao18L2, Kunst18, Liu19L, Lee19B, Edvardsson19, Lee19L, Ezawa19B1, Ezawa19B2, Yokomizo19, Song19L1, Song19L2, Longhi19, Okuma19,
	Okuma20, Borgnia20, Kawabata20, Hofmann20, Zhang19a, Yang19a, Yoshida19a, Yi20a, Liu20a, Helbig20, Xiao20, McDonald18, Longhi20, Yu20, Li20}.

Higher-order topological phases are classified into two types: \textit{intrinsic} and \textit{extrinsic}.
Intrinsic higher-order topology can be determined by bulk band structures.
Namely, intrinsic higher-order topological boundary modes are stable as long as the bulk gap is open 
\cite{Langbehn17, Song17, Schindler18, Khalaf18X, Khalaf18B, Ahn18, Matsugatani18, Geier18, Luka19, Takahashi20}.
If the symmetry protecting higher-order topology is preserved, the topological characterization is independent of crystal termination. 
By contrast, extrsinsic higher-order topology depends on crystal termination.
Thus, when the surfaces and edges close the band gap, extrinsic higher-order topology can change unlike the intrinsic one
\cite{Benalcazar17S, Benalcazar17B, Lin18, Geier18, Luka19, Hayashi18, Hayashi19, Okugawa19B}.
Accordingly, extrinsic higher-order topological corner modes generally stem from nontriviality in the edges
\cite{Langbehn17, Geier18, Luka19, Hayashi18, Hayashi19}.

In non-Hermitian systems, complex spectra provide a new topological classification based on a point gap
\cite{Gong18, Zhou19B, Liu19B, Kawabata19X, Kawabata19L, Okuma20, Borgnia20}.
Point-gap topology indicates how a non-Hermitian Hamiltonian can be deformed to a unitary matrix.
Thus, topological phenomena characterized by a point gap are unique to non-Hermitian systems.
Interestingly, the non-Hermitian skin effect is one of the point-gap topological examples \cite{Gong18, Ezawa19B1, Okuma20, Borgnia20, Zhang19a, Yang19a, Yoshida19a}.

The skin effect is a remarkable difference between the energy spectra under a periodic boundary condition (PBC) and those under an open boundary condition (OBC).
For a large open system, we can analyze the non-Hermitian energy spectrum by complex wavevectors $k$, according to the non-Bloch band theory \cite{Yao18L1, Yokomizo19}.
In one-dimensional (1D) systems, the value of $\beta\equiv{\rm e}^{ik}$ is confined on a loop on a complex plane.
The loop is called generalized Brillouin zone.
Because non-Hermitian eigenstates can have complex wavevectors,
the eigenstates called skin modes are localized at either end of the system under the OBC \cite{Yao18L1}. 
Intriguingly, non-Hermitian skin effects lead to many novel phenomena including nonreciprocal transport \cite{Song19L1,McDonald18,Longhi20,Yu20,Li20}.
Therefore, it is important to search for nontrivial skin effects for non-Hermitian physics.

There are two approaches to detect non-Hermitian skin effects.
As mentioned above, one is to directly compute the generalized Brillouin zone to describe the localization 
\cite{Yao18L1, Yao18L2, Liu19L, Yokomizo19, Song19L2, Zhang19a, Kawabata20, Yang19a}. 
However, it is difficult, especially in high-dimensional systems with many bands.
Another is to employ a topological invariant for point-gap topology \cite{Gong18, Ezawa19B1, Okuma20, Zhang19a, Yang19a, Yoshida19a}.
A nontrivial invariant for a point gap typically reveals that a skin effect has occurred.
Because point-gap topology gives a simple understanding of skin effects,
the approach can be extended to various systems.
Hence, we take the topological approach to generally explore skin effects.

In this Rapid Communication, we suggest skin effects characterized by second-order topology.
Our study connects non-Hermitian point-gap topology and Hermitian higher-order topology.
We study skin modes localized at the corners by using intrinsic and extrinsic second-order topology.
The work opens the door to discover non-Hermitian topological skin effects with higher-order topology.

\textit{Topology of skin effects.}
Because skin effects originate from point-gap topology,
we introduce the topological characterization of non-Hermitian systems \cite{Gong18, Kawabata19X, Okuma20}. 
A point gap for a Hamiltonian $H$ is open if $\det (H-E) \neq 0$ at a reference energy $E$. 
Point-gap topology for a non-Hermitian Hamiltonian $H$ at $E$ is given from the extended Hermitian Hamiltonian defined as
\begin{align}
	\tilde{H}=
	\begin{pmatrix}
		0 & H-E \\
		H^{\dagger}-E^{*} & 0
	\end{pmatrix}. \label{pointgap}
\end{align}
Importantly, the point-gap topology for $H$ coincides with Hermitian topology for the extended Hermitian Hamiltonian $\tilde{H}$ \cite{Gong18, Kawabata19X}. 
The extended Hermitian Hamiltonian has diagonal chiral symmetry represented as $\Gamma = \mathrm{diag}(1,-1)$.
Therefore, 1D non-Hermitian systems without any symmetries can be classified by a winding number for 1D Hermitian systems with chiral symmetry.
The integer winding number for the PBC is \cite{Gong18, Kawabata19X, Okuma20, Zhang19a, Yang19a}
\begin{eqnarray}
	W(E)=\int _0^{2\pi}\frac{dk}{2\pi i}\frac{d}{dk}\log \det (H(k)-E). \label{winding}
\end{eqnarray}
If there is a reference energy with nonzero $W(E)$,
a non-Hermitian skin effect occurs \cite{Okuma20, Zhang19a}. 
The topological origin is understandable as zero modes in the chiral-symmetric Hermitian Hamiltonian with a nonzero winding number \cite{Okuma20}.


We extend the idea to second-order topology with chiral symmetry by the winding number for skin effects.
To do so, we study the topological coincidence
between a non-Hermitian Hamiltonian and an extended Hermitian Hamiltonian with second-order topology.
This work reveals that intrinsic second-order topology protected by inversion symmetry can be characterized by a winding number for a skin effect.
We also investigate extrinsic second-order topology with nonzero winding numbers of the edge spectra.
Then, we show that second-order topology yields skin modes localized at the corners.


\textit{Intrinsic second-order topological skin effect.}
First, we investigate the skin effect with intrinsic second-order topology characterized by inversion symmetry.
When a two-dimensional (2D) chiral-symmetric Hermitian system has inversion symmetry,
intrinsic second-order topology is allowed if the operators of the two symmetries anticommute \cite{Khalaf18B, Matsugatani18}.
The topological zero-energy corner modes are protected by the bulk topology.
By using the second-order topology for a point gap,
we show that non-Hermitian systems can have skin modes localized at the corner. 

Before going into details of the skin effect, we explain the bulk-corner correspondence of the intrinsic second-order topology in Hermitian systems.
We assume that the Hermitian Hamiltonian $\tilde{H}$ has inversion symmetry $\tilde{I}$ which gives $\tilde{I}\tilde{H}(\bm{k}){\tilde{I}}^{\dagger}=\tilde{H}(-\bm{k})$
in addition to chiral symmetry $\Gamma$.
Such second-order topology is characterized by double band inversion,
which means that band touchings invert two conduction and two valence bands with the opposite parity eigenvalues
\cite{Khalaf18B, Khalaf18X, Ahn18, Matsugatani18, Wang19L, Park19, Lee20, Takahashi20, Ono18}.
The double band inversion induces topological bulk structures with intrinsic second-order topology for the corner modes. 
To relate the double band inversion to the winding number in Eq.~(\ref{winding}), we introduce the following bulk topological invariants defined as \cite{supplement}
\begin{align}
	\mu _{j=x,y}= n_{-}(0,0)-n_{-}(\pi, \pi)+ \tilde{s}_j\{n_{-}(\pi, 0)-n_{-}(0,\pi ) \},
\end{align}
where $n _{-}(\Gamma _i) $ is the number of states with odd parity below zero energy at the inversion-invariant momentum $\Gamma _i$,
and $\tilde{s}_{x(y)}=+1(-1)$.
The parity invariant $\mu _{x(y)}$ can detect the winding number for the ribbon geometry open in the $x (y)$ direction 
if the ribbon retains inversion symmetry.
Significantly, $\mu _j/2$ is equal to the winding number for the 1D ribbon geometry modulo 2
when the bulk is not first-order topological.
Namely, the winding number can become nonzero if $\mu _j\equiv 2 \pmod 4$.
Hence, the 2D system hosts zero modes under the full OBC, i.e., the topological corner states,
owing to the nonzero winding number.

From the above discussion, when a non-Hermitian Hamiltonian $H$ gives an extended Hermitian Hamiltonian $\tilde{H}$ with a nontrivial parity invariant,
$H$ has a nonzero winding number under the OBC. 
Thus, the non-Hermitian system exhibits skin modes localized at the corner.
Hereafter, we demonstrate a skin effect with intrinsic second-order topology.
We study the following 2D Hamiltonian under the PBC,
\begin{align}
	H_{\mathrm{int}}(\bm{k})&=(m-c\sum _{j=x,y}\cos k_j)s_0 + it\sin k_ys_x + it \sin k_x s_y \notag \\
	&-B_xs_x -B_y s_y, \label{NHint}
\end{align}
where $t, c, m$ and $\bm{B}=(B_x,B_y)$ are real parameters.
For simplicity, we set all the parameters positive.
Here, $s_{x,y,z}$ are Pauli matrices, and $s_0 $ is the identity matrix.
The extended Hermitian Hamiltonian is 
\begin{align}
	\tilde{H}_{int}(\bm{k})&=
	(m-\mathrm{Re}E -c\sum _{j=x,y}\cos k_j)\sigma _x\otimes s_0 \notag \\
	&-t\sin k_y \sigma _y\otimes s_x - t\sin k_x \sigma _y \otimes s_y \notag \\
	&-B_x\sigma _x\otimes s_x -B_y\sigma _x\otimes s_y 
	+ (\mathrm{Im}E)\sigma _y \otimes s_0, \label{Hermitianint}
	\end{align}
where $\sigma _{x,y,z}$ are Pauli matrices, and $\sigma _0 $ is the identity matrix.
Only when the reference energy $E$ is real does the Hermitian Hamiltonian have inversion symmetry represented by $\tilde{I}=\sigma _x \otimes s_0$.
Because $n_{-}(0,\pi ) = n_{-}(\pi,0)$ in this model, we put $\mu =\mu _x=\mu _y$.
Thus, we can search for $E$ with a nonzero winding number with the invariant $\mu$, independently of which direction is open.
Although $E$ can deviate from the real axis, the winding number does not change unless the point gap closes.

\begin{figure}[t]
	\includegraphics[width=8.5cm]{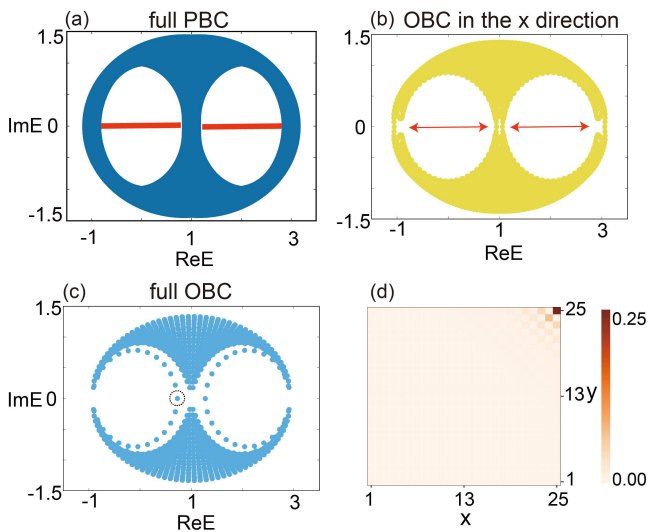}
	\caption{\label{intfig} (a)-(c) Energy spectra of the non-Hermitian Hamiltonian $H_{\mathrm{int}}$ under the PBC, the OBC only in the $x$ direction, and the full OBC.
		In (a), the red solid lines on the real axis indicate $(m-2c-|\bm{B}|, m-|\bm{B}|)$ and $(m+|\bm{B}|, m+2c-|\bm{B}|)$ with $\mu =-2$. 
		The gapped energy areas including the red arrows have the nonzero winding number at each of the reference points in (b). 
		(c) shows second-order topological skin modes.
		(d) Amplitude of the skin mode in the dotted circle in (c). The axes $x$ and $y$ represent the coordinates.
	}
\end{figure}

We assume that $|\bm{B}|$ is sufficiently smaller than the other parameters.
Then, the model under the PBC can have $\mu (E) =-2$ and trivial first-order topology
when the real $E$ is in the region 
$(m-2c-|\bm{B}|, m-|\bm{B}|)$ or $(m+|\bm{B}|, m+2c-|\bm{B}|)$.
Details are given in the Supplemental Material \cite{supplement}.
Thus, the model has a nonzero winding number in the inversion-symmetric ribbon geometry
with one direction open.
We put $m=t=c=1$ and $B_x=B_y=0.1$ to calcuate the second-order topological skin modes.
Figure \ref{intfig} shows the energy spectrum under various boundary conditions.
In contrast to the spectrum under the PBC and the ribbon geometry in Figs.~\ref{intfig} (a) and \ref{intfig} (b),
skin modes are widely distributed under the full OBC in Fig.~\ref{intfig} (c).
We can see that the skin effect occurs around the energy region with $\mu (E)=-2$.
As seen in Fig.~\ref{intfig}(d), the skin modes are localized at the corner.
The skin modes are localized at one corner in each of the areas with the intrinsic second-order topology.

Here, we discuss inversion symmetry in an extended Hermitian Hamiltonian.
The original non-Hermitian Hamiltonian $H_{\mathrm{int}}(\bm{k})$ in Eq.~(\ref{NHint}) is not inversion-symmetric.
However, $H_{\mathrm{int}}(\bm{k})$ satisfies $H_{\mathrm{int}}(\bm{k})=U_IH_{\mathrm{int}}^{\dagger}(-\bm{k})U_I^{\dagger}$
with the unitary matrix $U_I=\sigma _0$ with $U_I^2=1$.
Then, the extended Hermitian Hamiltonian with a real reference energy obtains inversion symmetry represented by
\begin{align}
	\tilde{I}=
	\begin{pmatrix}
		0 & U_I \\
		U_I & 0
	\end{pmatrix}, \label{inversionmat}
	\end{align}
where $\tilde{I}^2=\tilde{I}\tilde{I}^{\dagger}=1$ and $\Gamma \tilde{I} \Gamma ^{-1}=-\tilde{I}$.
The inversion operator anticommuting with the chiral operator can yield intrinsic second-order topology \cite{Khalaf18B, Matsugatani18}.
The above conditions generally give inversion symmetry to extended Hermitian Hamiltonians.

\textit{Extrinsic second-order topological skin effect.}
Next, we study the skin effect whose origin is extrinsic second-order topology characterized only by chiral symmetry.
In the extrinsic higher-order topological phase,
the edges are topologically nontrivial while the bulk is trivial \cite{Geier18, Langbehn17, Luka19, Hayashi18, Hayashi19, Okugawa19B}.
In other words, the edge bands are topological in view of the 1D first-order topology from chiral symmetry.
The edge-corner correspondence produces topological corner modes through the edge gap closing at zero energy.
Thus, the nontrivial topology on the 1D edges can be used for skin modes localized at the corners.

To see the skin effect with extrinsic higher-order topology,
we use a 2D second-order topological Hermitian Hamiltonian whose topological edge gap closing is analytically obtainable \cite{Hayashi18, Hayashi19, Okugawa19B}.
The Hamiltonian is 
\begin{align}
\tilde{H}_{ext}'(\bm{k})=\mathcal{H}_x(k_x)\otimes 1_y + \Pi _x \otimes \mathcal{H}_y(k_y), \label{geneHext}
\end{align}
where $\mathcal{H}_x(k_x)$ and $\mathcal{H}_y(k_y)$ are Hermitian matrices with chiral symmetry which we write as $\Pi _x$ and $\Pi _y$, respectively.
$1_y$ is the identity matrix with the same size as $\mathcal{H}_y$.
Then, $\tilde{H}_{ext}'(\bm{k})$ has chiral symmetry $\Pi = \Pi _x \otimes \Pi _y$.
When we impose a semi-infinite OBC with a right-angled corner,
the number of zero-energy corner states are characterized by a $\mathbb{Z}$ topological invariant given by \cite{Hayashi18, Hayashi19, Okugawa19B}
\begin{align}
	\nu _{2D}=w_xw_y, \label{topoext}
	\end{align}
where $w_{i=x,y}$ are conventional winding numbers for the chiral-symmetric Hermitian matrices $\mathcal{H}_i(k_i)$. 
The topological invariant $\nu _{2D}$ changes when the edges close the gap at zero energy as well as the bulk \cite{Hayashi18, Hayashi19, Okugawa19B}.
Therefore, the edge gap closing gives nonzero winding numbers for the edge bands whose topology corresponds to the extrinsic second-order topology \cite{supplement}.

When a non-Hermitian Hamiltonian gives an extended Hermitian Hamiltonian with  extrinsic second-order topology,
the edge nontriviality gives rise to skin modes localized at the corner.
To confirm our theory,
we study the following 2D non-Hermitian Hamiltonian,
\begin{align}
H_{\mathrm{ext}}(\bm{k})&=2t_x\cos k_x\tau _0 \notag \\
&-2ig_y\sin k_y\tau _x -2it_y\cos k_y\tau _y -2ig_x\sin k_x\tau _z, \label{Hextk}
\end{align}
where $t_{x,y}$ and $g_{x,y}$ are real parameters.
We assume that $t_x$ and $t_y$ are positive for simplicity.
$\tau _{x,y,z}$ are Pauli matrices, and $\tau _0$ is the identity matrix.
The extended Hermitian Hamiltonian with $H_{\mathrm{ext}}$ can show the second-order topology given by $\nu _{2D}$ in Eq.~(\ref{topoext}).

We impose OBCs with parallel edges on the model to calculate the winding numbers.
We study the real $E$ since the winding number is unchanged if the point gap remains open. 
In this model, $[w_x(E), w_y(E)]=[-\mathrm{sgn}(g_x),-\mathrm{sgn}(g_y)]$ when $|E|<2t_x$, and otherwise $[w_x(E), w_y(E)]=[0,-\mathrm{sgn}(g_y)]$.
Therefore, the energy real axis in $|E|<2t_x$ has the extrinsic second-order topology with $\nu _{2D}(E)=\mathrm{sgn}(g_xg_y)$ because of the nontrivial edges.
When the topology changes on the real axis,
the point gap closes at $E=\pm 2t_x$ on the edges open in the $y$ direction but periodic in the $x$ direction.
We analyze the second-order topology by studying the change of $\nu _{2D}$ in the Supplemental Material \cite{supplement}.

\begin{figure}[t]
	\includegraphics[width=8.5cm]{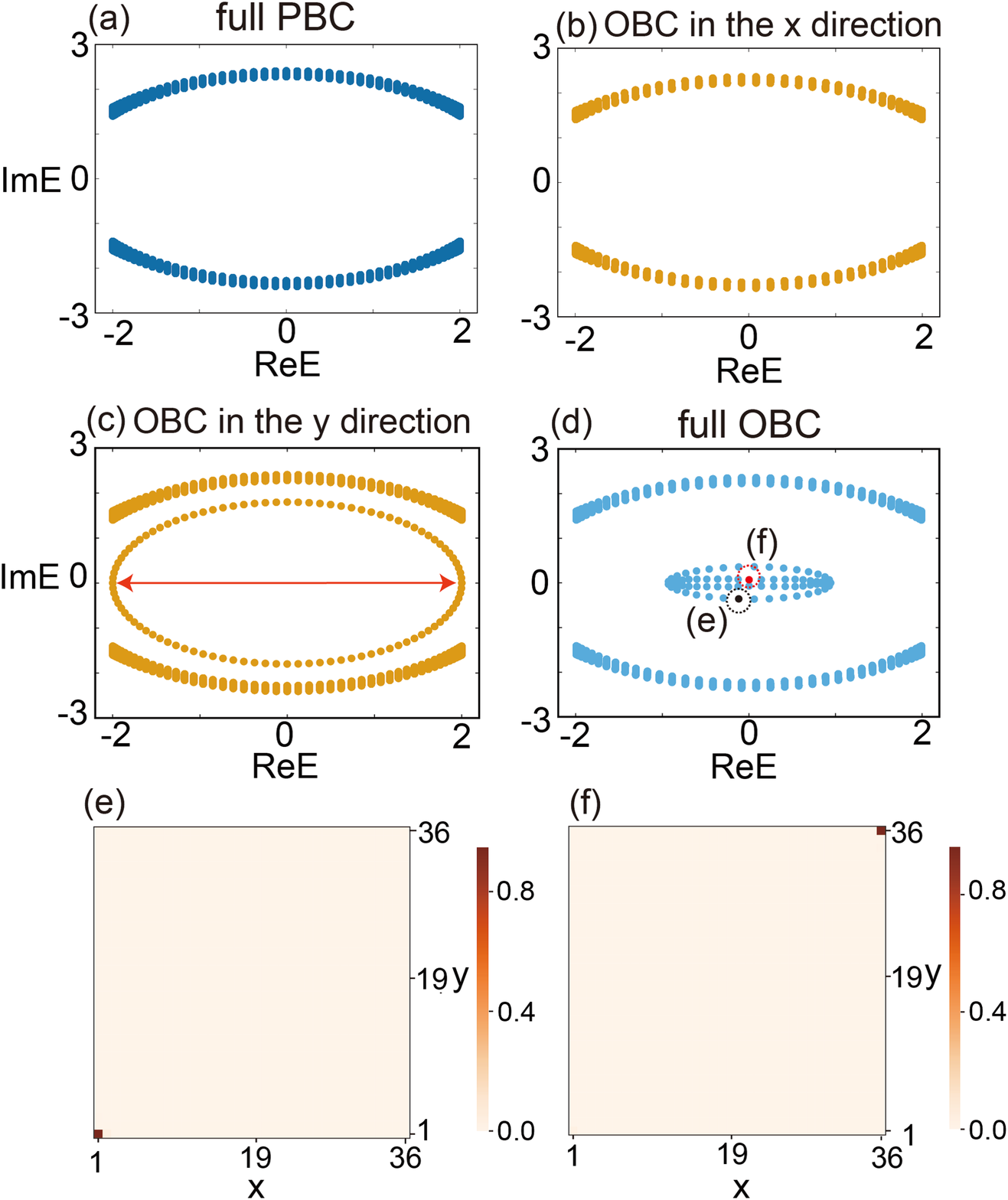}
	\caption{\label{extfig} Energy spectra of the non-Hermitian Hamiltonian $H_{\mathrm{ext}}$
		under (a) the full PBC, (b) the OBC only in the $x$ direction, (c) the OBC only in the $y$ direction, and (d) the full OBC.
		The ellipse in (c) is the edge spectrum under the OBC in the $y$ direction.
		Each spectrum of the two $x$-periodic edges gives the nonzero winding number.
		The red arrow in (c) indicates $|E|<2t_x$ with the finite $\nu _{2D}$ for the nonzero winding numbers inside the edge spectra.
		In (d),the skin modes can be found in the energy region inside the edge spectra in (c).
		(e) and (f) Amplitude of the black and the red skin modes in the dotted circles in (d). 
		Each of the skin modes are localized at the opposite corners because the two edge spectra contribute to the skin effect.
	}
\end{figure}

We compute the energy spectra of the non-Hermitian Hamiltonian $H_{\mathrm{ext}}$ under different boundary conditions.
We set the parameters $(t_x, g_x, t_y, g_y)=(1.0, 0.9, 0.8, 0.7)$ for the calculation.
The model in Eq.~(\ref{Hextk}) has trivial point-gap topology under the full PBC, as shown in Fig.~\ref{extfig} (a).
Thus, a skin effect does not occur when we impose the OBC either in the $x$ or in the $y$ direction [Figs.~\ref{extfig} (b) and \ref{extfig}(c)].
However, Fig.~\ref{extfig} (c) shows that edge spectra give nonzero winding numbers under the OBC in the $y$ direction.
We emphasize that the winding structures consist of the spectra of the two edges parallel to the $x$ direction.
In Fig.~\ref{extfig}(d), we can see the skin modes under the full OBC near the region with the nonzero $\nu _{2D}(E)$.
Because the system has the two nontrivial edge bands, each of the edges contributes to the skin effect.
Hence, we can find that the two edges give rise to the skin modes localized at the different corners [Figs.~\ref{extfig} (e) and \ref{extfig}(f)],
unlike intrinsic second-order topological skin modes.

\textit{Conclusion and discussion.}
In this Rapid Communication, we have proposed non-Hermitian skin modes localized at the corners
by the second-order topology of chiral-symmetric Hermitian Hamiltonians.
We have elucidated the intrinsic and extrinsic second-order topological skin effects.
The instrinsic second-order topological skin effect can be detected by inversion symmetry.
The extrinsic second-order topological skin effect can be understood from the edge-corner correspondence.
By using our theory, we can generally construct non-Hermitian Hamiltonians for the second-order topological skin effects.

Some works also discussed skin modes localized at the corners under a full open boundary condition 
\cite{Lee19L, Ezawa19B1}.
The skin effects in previous works were studied by bulk first-order topology \cite{comment}.
In comparison with previous works, we have predicted skin effects characterized only by second-order topology.
Moreover, since our theory defines the second-order topological invariants for each reference energy,
they allow us to easily find the skin modes under full open boundary conditions.

For both the intrinsic and the extrinsic cases, 
we can understand the nontrivial skin effects by connecting the second-order topology to the conventional topology for one-dimensional chiral-symmetric systems.
However, the topological structures are different.
Intrinsic second-order topology stems from a nontrivial energy spectrum under a full periodic boundary condition.
As a result, a ribbon geometry open in one direction has a nonzero winding number which leads to skin modes.
Meanwhile, for extrinsic higher-order topology, the point-gap topology of the energy spectrum is trivial under the full periodic boundary condition.
When one direction is open, the system shows nontrivial topology at each of the edges.
Therefore, intrinsic and extrinsic second-order topological skin modes are differently localized at corners. 
Furthermore, because point-gap topology reflects higher-order topology of extended Hermitian Hamiltonians,
the results will be generalized to various skin effects due to higher-order topology characterized by different symmetries.

\textit{Note added.}
Recently, we became aware of several related works about second-order topological skin effects \cite{Denner20, Kawabata20a, Fu20}.

We thank S. Hayashi and K. Ohsumi for fruitful discussions.
We also thank Y. Fu and C. H. Lee for valuable discussions on their related works.
This work was supported by JSPS Grant-in-Aid for Scientific Research on Innovative Areas "Discrete Geometric Analysis for Materials Design" Grant No. JP17H06469,
and JSPS KAKENHI (Grants No. JP18J22113 and No. JP18J23289).

\bibliographystyle{apsrev4-1}
\bibliography{skin}

\begin{thebibliography}{18}
\bibitem [{com()}]{commentsup}%
  \BibitemOpen
  \href@noop {} {\bibinfo  {journal} {We do not consider a phase transition
  from a trivial phase with $[w_x, w_y]=[0,0]$ to a topological phase with
  $[-1,-1]$ because the phase transtion happens from the bulk gap closing}\
  }\BibitemShut {NoStop}%
\end{thebibliography}

\end{document}



\title{Supplemental material for "Second-order topological non-Hermitian skin effects"
}

\author{Ryo Okugawa}
\affiliation{%
	Graduate School of Information Sciences, Tohoku University, Sendai 980-8579, Japan
}%
\author{Ryo Takahashi}
\affiliation{%
	Department of Physics, Tokyo Institute of Technology, 2-12-1 Ookayama, Meguro-ku, Tokyo, 152-8551, Japan
}%

\author{Kazuki Yokomizo}
\affiliation{%
	Department of Physics, Tokyo Institute of Technology, 2-12-1 Ookayama, Meguro-ku, Tokyo, 152-8551, Japan
}%

\renewcommand{\thefigure}{S\arabic{figure}}
\renewcommand{\theequation}{S\arabic{equation}}

\maketitle

\section{Intrinsic second-order topology protected by inversion symmetry}
\subsection{The relationship between the bulk parity invariant and the winding number for ribbon geometry} \label{secparity}

To obtain skin modes localized only at corners in 2D systems,
we need to calculate the winding number under the system which is open in one direction.
In this section, we reveal a relationship between the winding number and the bulk pariy invariant $\mu _j$
for extended Hermitian Hamiltonians with inversion symmetry.
When the $x_j$ direction is open, we show that the winding number is given by
\begin{align}
	W_{x_j-OBC}=\frac{\mu _j}{2} \pmod 2. \label{Wribbon}
	\end{align}
Therefore, if $\mu _j\equiv 2 \pmod 4$, the correspondent non-Hermitian Hamiltonian shows skin modes localized at the corner.

We begin with the winding number for 1D chiral-symmetric Hermitian systems in the presence of inversion symmetry.
The winding number can be expressed by the number of negative parity eigenvalues below zero energy at $k=0$ and $\pi$.
The winding number is rewritten as \cite{Mondragon14,Hughes11, Turner12, Benalcazar17B}
\begin{align}
	(-1)^W=(-1)^{n_{-}(0)-n_{-}(\pi)}, \label{Winv}
	\end{align}
where $n_{-}(0)$ and $n_{-}(\pi)$ are the number of negative parity eigenvalues below zero energy at $k=0$ and $\pi$, respectively.
In 2D systems with inversion and chiral symmetries,
the ribbon geometry with the $x_j$ direction open can be regarded as 1D.
Thus, we can obtain the winding number $W_{x_j-OBC}$ by applying Eq.~(\ref{Winv}) to the ribbon geometry.

Under the OBC in the $x_j$ direction in the 2D system,
we denote the number of the negative parity eigenvalues below zero energy at $k=0 ~(\pi)$ as $\tilde{n}_{-}^{x_j}(0) ~(\tilde{n}_{-}^{x_j}(\pi) )$.
Here, we assume that the bands are gapped at zero energy both under the PBC and under the OBC in the $x_j$ direction. 
Then, the band structures do not have the nontrivial first-order topology on the assumption.
According to Ref.~\onlinecite{Takahashi20}, 
the parity invariants $\tilde{n}_{-}^{x_j}(0)$ and $\tilde{n}_{-}^{x_j}(\pi)$ can be obtained via cutting procedure which relates the PBC to the OBC \cite{Teo08}.
From the procedure, $\tilde{n}_{-}^{x_j}(0)$ and $\tilde{n}_{-}^{x_j}(\pi)$ are related 
to the number of the 2D negative parity eigenvalues under the PBC.
The relationships are \cite{Takahashi20}
\begin{align}
	&\tilde{n}^{x}_{-}(0)=\frac{N_{OBC}^x(0)-n_{PBC}}{2}+\frac{n_{-}(0,0)+n_{-}(\pi, 0)}{2}, \\
	&\tilde{n}^{x}_{-}(\pi )=\frac{N_{OBC}^x(\pi)-n_{PBC}}{2}+\frac{n_{-}(0,\pi )+n_{-}(\pi ,\pi )}{2}, \\
	&\tilde{n}^{y}_{-}(0)=\frac{N_{OBC}^y(0)-n_{PBC}}{2}+\frac{n_{-}(0,0)+n_{-}(0, \pi )}{2}, \\
	&\tilde{n}^{y}_{-}(\pi )=\frac{N_{OBC}^y(\pi )-n_{PBC}}{2}+\frac{n_{-}(\pi, 0)+n_{-}(\pi ,\pi )}{2}, 
	\end{align}
where $N_{OBC}^{x_j}(0)$ and $N_{OBC}^{x_j}(\pi )$ are the number of occupied states under the OBC in the $x_j$ direction,
and $n_{PBC}$ is the number of occupied bands under the PBC. 
We have $N_{OBC}^{x_j}(0) = N_{OBC}^{x_j}(\pi )$ since the bands are assumed to be gapped at zero energy under the OBC.
Therefore, we obtain
\begin{align}
	\tilde{n}^{x_j}_{-}(0)-\tilde{n}^{x_j}_{-}(\pi )=\frac{\mu _j}{2}. \label{ndif}
	\end{align} 
As a result, Eq.~(\ref{Wribbon}) is proved by Eqs.~(\ref{Winv}) and (\ref{ndif}).
Because the winding number for corner modes can be determined by the bulk parity invariant, the second-order topology is intrinsic. 
Furthermore, we here show that $W_{x_j-OBC}$ can be defined only if $\mu_ j$ is an even integer.
We have $(-1)^{\mu _x}=(-1)^{\mu _y}=(-1)^{\sum _{i}n_{-}(\Gamma _i)}$, where the sum runs over all the TRIM $\Gamma _i$ in the 2D bulk momentum space.
Meanwhile, if $(-1)^{\sum _{i}n_{-}(\Gamma _i)}=-1$, the 2D chiral-symmetric system is gapless under the PBC \cite{Ono18}.
Hence, we can apply $W_{x_j-OBC}$ when $\mu _j$ is an even integer.

\begin{figure}[t]
	\includegraphics[width=10cm]{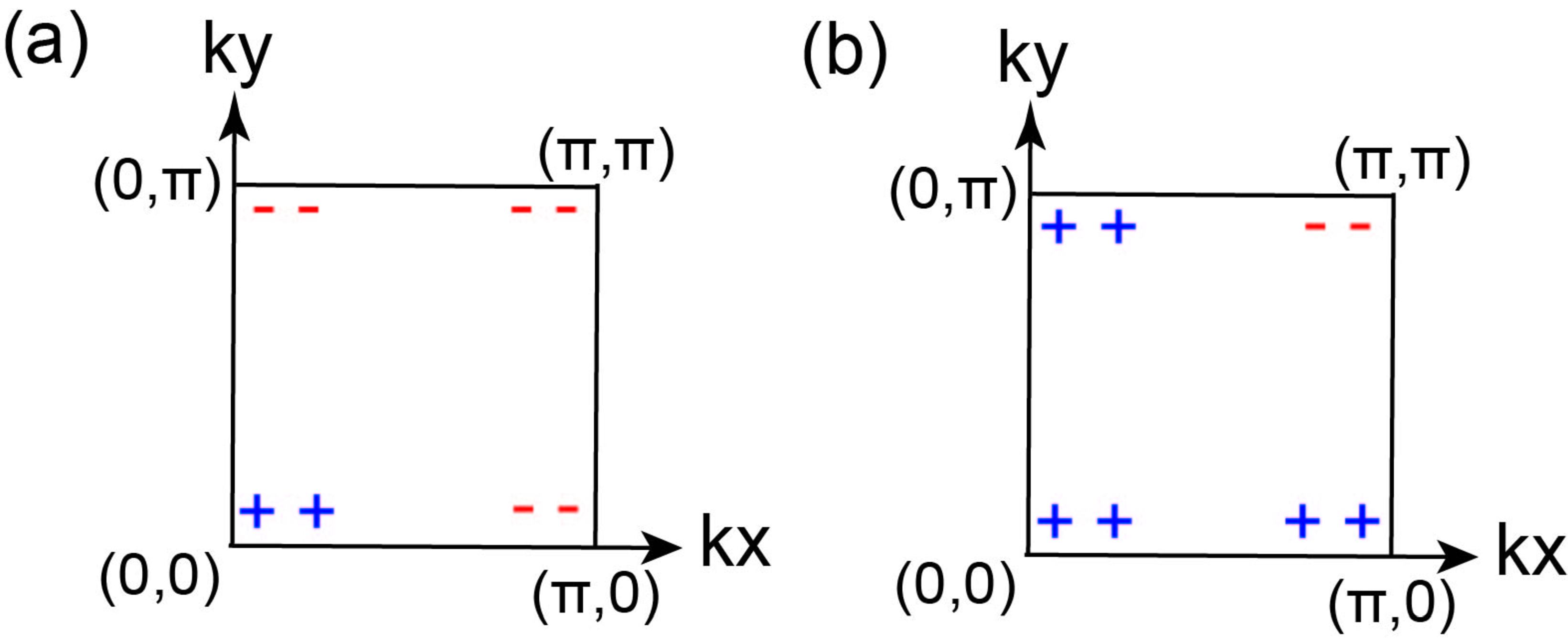}
	\caption{\label{parityfig} Illustrative examples of parity eigenvalues below zero energy to give $\mu =-2$. 
		The signs $+ (-)$ indicate the parity eigenvalues below zero energy at the inversion-invariant momenta.
		In these cases, the system is not first-order topological.
		When $m-2c-|\bm{B}|<E<m-|\bm{B}|$ and $m+|\bm{B}|<E<m+2c-|\bm{B}|$, the model $\tilde{H}_{\mathrm{int}}(\bm{k})$ realizes (a) and (b), respectively.
	}
\end{figure}

Figure \ref{parityfig} illustrates examples for $\mu _x= \mu _y =- 2$.
The extended Hermitian Hamiltonian can have $\mu _j\equiv 2 \pmod 4$
when double band inversion occurs at an odd number of the inversion-invariant momenta.
Inversion symmetry needs to anticommute with chiral symmetry to change $\mu _i$ through double band inversion \cite{Khalaf18B}.


Henceforth, we consider how to realize systems with $\mu _j\equiv 2 \pmod 4$.
To do so, we use a non-Hermitian Hamiltonian $H$ which gives time-reversal symmetry $\tilde{T}$ with $\tilde{T}^2=-1$
to the extended Hermitian Hamiltonian $\tilde{H}$.
In this case, we can compute the $\mathbb{Z}_2$ Kane-Mele topological invariant under the PBC \cite{Kane05, Fu06, Fu07}.
If $H$ also gives inversion symmetry $\tilde{I}$ to $\tilde{H}$ from Eq.~(6),
the $\mathbb{Z}_2$ invariant can be calculated from the number of negative parity eigenvalues at time-reversal invariant momenta (TRIM) below zero energy \cite{Fu07}.
Because of Kramers degeneracy, the $\mathbb{Z}_2$ invariant is given by
\begin{align}
	(-1)^{\nu}=(-1)^{\sum _{i}n_{-}(\Gamma _i)/2}. \label{nuz2}
	\end{align}
The sum is taken over all the TRIM in the 2D bulk momentum space.
Therefore, when $\tilde{H}$ shows bulk parity eigenvalues in Figs.~\ref{parityfig} (a) and (b),
the $\mathbb{Z}_2$ topology is nontrivial.
Then, if we slightly break only the time-reversal symmetry in the $\mathbb{Z}_2$ nontrivial case,
we realize double band inversion for $\mu _j\equiv 2 \pmod 4$ as long as the band gap is open.

Finally, we mention a 2D non-Hermitian skin effect studied in Ref.~\onlinecite{Okuma20}.
The skin effect is protected by first-order topology from time-reversal symmetry in class $\mathrm{AII}^{\dagger}$.
The skin modes appear under OBC only in one direction, whereas they vanish under full OBC since the skin effect stems from the first-order topology.
The previous work showed that
a topological defect is necessary to obtain the skin modes under full OBC.
Then, the skin modes are localized near the boundary and the defect.
In comparison with the previous work,
our case breaks time-reversal symmetry, but keeps inversion symmetry for the second-order topology in the extended Hermitian Hamiltonian.
The skin modes are characterized by the intrinsic second-order topology,
and therefore they are localized at corners under full OBC.

\subsection{Calculation of the nontrivial parity invariant $\mu$ for the skin modes in the model}
By using the discussion in Sec.~\ref{secparity},
we show how to obtain $\mu _{i=x,y}(E)=-2$ for $\tilde{H}_{\mathrm{int}}$ in Eq.~(5) in the main text.
The nontrivial parity invariants yield the intrinsic second-order topological skin effect for $H_{\mathrm{int}}$ in Eq.~(4) in the main text.
We here calculate parity eigenvalues for the extended Hermitian Hamiltonian with the real reference energy. 

To realize $\mu _j\equiv 2 \pmod 4$, we ignore the parameter $\bm{B}$ at first.
When $\bm{B}=0$ and $E$ is real, the extended Hermitian Hamiltonian in Eq.~(5) has not only inverson symmetry but also time-reversal symmetry.
The time-reversal symmetry is represented by $\tilde{T}=-i\sigma _x\otimes s_yK$, where $K$ is complex conjugation.
Therefore, Kramers degeneracy arises because $\tilde{T}^2=-1$.
At all the TRIM $\Gamma _i$, we can analytically obtain parity eigenvalues $\xi (\Gamma _i)$
for the Kramers pair below zero energy.
By putting $h(\bm{k})=m-E-c\sum _{j=x,y}\cos k_j$, the parity eigenvalues are 
\begin{align}
	\xi (\Gamma _i)=-\mathrm{sgn}(h(\Gamma _i)). \label{parity}
\end{align}
In this model, $n_-(\Gamma _i)=2 (0)$ if $\xi (\Gamma _i)$ is negative (positive) since Kramers degeneracy is present.
For the time-reversal symmetric model, we can easily obtain the $\mathbb{Z}_2$ Kane-Mele topological invariant $\nu$.
From Eqs.~(\ref{nuz2}) and (\ref{parity}),
the topological invariant $\nu$ is obtained from 
\begin{align}
	(-1)^{\nu (E)}=\mathrm{sgn}(m-E-2c)\mathrm{sgn}(m-E+2c).
\end{align}
Because $m$ and $c$ are positive in the model,
$\nu (E) =1$ , i.e. $\mu (E)\equiv 2 \pmod 4$ when $m-2c<E<m+2c$ except $E=m$.
$h(0, \pi) = h(\pi ,0)$ leads to $n_{-}(0, \pi) = n_{-}(\pi, 0)$, and therefore we have $\mu _x = \mu _y =\mu $.
If an odd number of the TRIM gives the negative parity eigenvalues, this model has $\nu =1$, and then $\mu =-2$.

We need to add the parameters $\bm{B}=(B_x, B_y)$ to $\tilde{H}_{\mathrm{int}}(\bm{k})$ in order to break the time-reversal symmetry.
Then, the invariant $\mu _j\equiv 2 \pmod 4$ detects the winding number for the inversion-symmetric 1D ribbon geometry.
As long as the bulk gap keeps open, the value of $\mu$ is preserved even though $\bm{B}$ is added.
As a result, we can obtain $\mu =-2$ in the absence of the time-reversal symmetry
if the finite $\bm{B}$ does not close the bulk gap.

For the extended Hermitian Hamiltonian $\tilde{H}_{\mathrm{int}}(\bm{k})$, the energy eigenvalues at $\Gamma _i$ are $\pm|h(\Gamma _i)|\pm |\bm{B}|$.
Therefore, when a reference energy $E$ satisfies $|h(\Gamma _i)|=|\bm{B}|$, $\mu (E)$ changes at the energy.
Hence, if $|\bm{B}|$ is sufficiently small, 
we obtain $\mu (E) =-2$ for the real $E$ in the region $m-2c-|\bm{B}|<E<m-|\bm{B}|$ and $m+|\bm{B}|<E<m+2c-|\bm{B}|$.
In the ribbon geometry, the system realizes the nonzero winding number near the regions
which give second-order topological skin modes [Fig.~1 (c)].
On the other hand, when $E<m-2c-|\bm{B}|$ or $E>m+2c-|\bm{B}|$,
the topology is trivial in terms of both the first order and the second-order topology.
The other regions of the real energy axis are gapless.

\section{Extrinsic second-order topology with only chiral symmetry}  
\subsection{Edge gap-closing and change of the topological invariant $\nu _{2D}$} \label{extrinsictopo}
We discuss how gap closing on edges changes the extrinsic second-order topology for $\tilde{H}_{\mathrm{ext}}'(\bm{k})$ in Eq.~(7) in the main text.
The topological invariant $\nu _{2D}=w_xw_y$ is defined for the matrix form in Eq.~(7).
The extrinsic second-order topology leads to edge bands with nontrivial winding structures.
In this section, we explain that change of $\nu _{2D}$ accompanies the edge gap-closing,
according to Ref.~[\onlinecite{Hayashi18, Hayashi19, Okugawa19B}].

Suppose that a phase transition happens from a trivial phase with $[w_x, w_y]=[0,-1]$ to a topological phase with $[-1,-1]$.
In this case, the band gap closes at zero energy on the edges periodic in the $x$ direction.
To see this, we consider the semi-infinite system with one edge periodic in the $x$ direction.
In other words, we impose the OBC in the $y$ direction on the system.
Since $\mathcal{H}_y$ is nontrivial thanks to $w_y=-1$, 
there is a zero mode $\phi _y^{zero}$ which satisfies $\mathcal{H}_y^{OBC}\phi _y^{zero}=0$.
Thus, $\tilde{H}'_{\mathrm{ext}}$ under the semi-infinite OBC has edge states represented by $\psi _n(k_x) \otimes \phi _y^{zero}$,
where $\psi _n(k_x)$ is the Bloch wavefunction for $\mathcal{H}_x(k_x)$.
Because $w_x$ changes the value from $0$ to $-1$,
any of the Bloch wavefunctions need to take the zero-valued eigenvalues.
As a result, the edge states $\psi _n(k_x) \otimes \phi _y^{zero}$ closes the gap at zero energy.
In a similar way, if the system enters a topological phase with $[-1,-1]$ from a trivial phase with $[-1,0]$,
gap closing occurs on the edges periodic in the $y$ direction.
Eventually, the edge states close the gap at zero energy when $\nu _{2D}$ changes,
which gives the nonzero winding number $W$ for the edge bands \cite{commentsup}, as shown in Fig.~2(c) in the main text.

Additionally, we can easily find the topological corner zero-energy modes in this model $\tilde{H}_{\mathrm{ext}}'$ as follows.
We assume that both $\mathcal{H}_x(k_x)$ and $\mathcal{H}_y(k_y)$ are topologically nontrivial, i.e., $\nu _{2D} \neq 0$.
Let us denote a zero mode as $\phi _i^{zero}$ for $\mathcal{H}_i$ under semi-infinite OBCs.
Then, the system shows a topological corner mode given by $\phi _x^{zero} \otimes \phi _y^{zero}$ under the semi-infinite OBC with one corner.
From this discussion, we can also see that the edges are nontrivial. 

\subsection{Calculation of the extrinsic second-order topology for the skin modes in the model}
The model $H_{\mathrm{ext}}(\bm{k})$ in Eq.~(9) in the main text shows skin modes localized at the corners.
To clarify the origin, we study extrinsic second-order topology of the extended Hermitian Hamiltonian.
Here, we show that the system has reference points with the nonzero $\nu _{2D}$ for the nonzero winding numbers on the edges,
which realizes the extrinsic second-order topological skin effect.

When the reference energy is real, we can easily show that the extended Hermitian Hamiltonian of $H_{\mathrm{ext}}(\bm{k})$ can have the extrinsic second-order topology.
The extended Hermitian Hamiltonian under the PBC is given by
\begin{align}
	\tilde{H}_{\mathrm{ext}}(\bm{k})=
	\begin{pmatrix}
		0 & H_{\mathrm{ext}}(\bm{k})-E \\
		H_{\mathrm{ext}}^{\dagger}(\bm{k})-E & 0
	\end{pmatrix}
\end{align}
In order to investigate the second-order topology, we perform unitary transformation for ${H}_{\mathrm{ext}}(\bm{k})$ represented as
\begin{align}
	U=
	\begin{pmatrix}
		0 & 0 & 0 & -1 \\
		1 & 0 & 0 & 0 \\
		0 & -1 & 0 & 0 \\
		0 & 0 & 1 & 0
	\end{pmatrix}.
\end{align}
Then, we obtain the following Hamiltonian:
\begin{align}
	\tilde{H}_{\mathrm{ext}}'(\bm{k})&=U\tilde{H}_{\mathrm{ext}}(\bm{k})U^{\dagger} \notag \\
	&=\mathcal{H}_x(k_x) \otimes 1_2 + \Pi \otimes \mathcal{H}_y(k_y). \label{Uex}
\end{align}
Here, $1_2$ is the $2 \times 2$ identity matrix, and $\Pi = \mathrm{diag}(1,-1)$.
In Eq.~(\ref{Uex}), $\mathcal{H}_{x,y}(k_{x,y})$ are
\begin{align}
	\mathcal{H}_{x}(k_x)
	&=\begin{pmatrix}
		0 & 2t_x\cos k_x -E -2ig_x\sin k_x \\
		2t_x\cos k_x -E +2ig_x\sin k_x& 0
	\end{pmatrix} \\
	\mathcal{H}_{y}(k_y)
	&=\begin{pmatrix}
		0 & 2t_y\cos k_y -2ig_y\sin k_y\\
		2t_y\cos k_y +2ig_y\sin k_y& 0 
	\end{pmatrix}.
\end{align}
Because $\mathcal{H}_x$ and $\mathcal{H}_y$ have chiral symmetry represented by $\mathrm{diag}(1,-1)$,
$\tilde{H}_{\mathrm{ext}}'(\bm{k})$ can be included in the Hamiltonians given by Eq.~(7).
$w_x$ and $w_y$ can be also calculated by Eq.~(2) in the main text.
The winding number for $\mathcal{H}_y$ are $w_y=-\mathrm{sgn}(g_y)$ since $t_y$ is positive now.
On the other hand, the winding number $w_x$ for $\mathcal{H}_x$ depends on the reference energy $E$.
Thus, while $w_x(E)=-\mathrm{sgn}(g_x)$ when $-2t_x<E<2t_x$,
the value becomes zero in the other regions on the real energy axis.
Hence, we obtain the nonzero topological invariant $\nu _{2D}(E) = \mathrm{sgn}(g_x)\mathrm{sgn}(g_y)$ if and only if $-2t_x<E<2t_x$.

As discussed in Sec.~\ref{extrinsictopo},
the Hermitian model in Eq.~(\ref{Uex}) shows topological phase transitions on the edges 
when the topological invariant $\nu _{2D}$ changes.
Then, the correspondent non-Hermitian Hamiltonian also closes the point gap at the reference energy $E$.
Indeed, if $[w_x, w_y]$ changes from $[0,-1]$ to $[-1,-1]$,
the gap closing needs to happen on edges along the $y$ direction \cite{Hayashi18, Hayashi19, Okugawa19B}.
Therefore, the point gap closes at $E=\pm 2t_x$ on the edges periodic in the $y$ direction, as seen in Fig.~2(c).
Consequently, the winding number can be finite on the edges.

We note that the origin of the edge spectra in this model is Hermitian topological edge states.
Let us focus on the $k_y$-dependent terms $h_y(k_y)=-2ig_y \sin k_y \tau _x -2it_x \cos k_y \tau _y$ in $H_{\mathrm{ext}}(\bm{k})$.
By multipling $h_y(k_y)$ by $i$, the terms become a 1D Hermitian topological model with the chiral symmetry $\tau _z$.
Thus, the terms produce the topological edge states under the OBC in the $y$ direction.
The edge states form the winding structures by the other terms, and therefore the second-order topological skin modes are realized under the full OBC. 
Hence, the extrinsic second-order topological skin modes can be also understood as skin-topological modes \cite{Lee19L, Fu20}.
However, extrinsic higher-order topological skin effects do not generally require topological edge states due to bulk topology
because perturbations to edges can give the nonzero winding number of the edge spectra, for instance. 


\bibliographystyle{apsrev4-1}
\bibliography{skin}